\documentclass[aps,prb,twocolumn,10pt,amsfonts,amssymb,longbibliography,superscriptaddress]{revtex4-2}
\usepackage{graphicx}
\usepackage{multirow}
\usepackage{epstopdf} 
\usepackage{dcolumn}
\usepackage{bm}
\usepackage{bbm,bbold}
\usepackage{color}
\usepackage{xcolor, soul}
\sethlcolor{green}
\usepackage{amsfonts}
\usepackage{amsmath}
\usepackage{mdframed}
\usepackage[normalem]{ulem}
\usepackage{mathrsfs}  
\usepackage[none]{hyphenat}
\usepackage{subfigure}
\usepackage{float}
\usepackage{physics}
\usepackage{amsmath,amssymb}
\usepackage[colorlinks=true,citecolor=blue]{hyperref}
\hypersetup{colorlinks=true,citecolor=blue,linkcolor=blue,urlcolor=blue}

\begin{document}
	\title{Skew scattering induced contribution to orbital Hall response}
  \author{Dayana Joy}
  \email{dayana_joy@srmap.edu.in}
  \affiliation{Department of Physics, School of Engineering and Sciences, SRM University AP, Amaravati, 522240, India}
  \author{Vivek Pandey}
  \email{vivek_pandey@srmap.edu.in}
  \email{vivek.p@iitb.ac.in}
  \affiliation{Department of Physics, Indian Institute of Technology Bombay, Powai, Mumbai 400076, India}
  \affiliation{Department of Physics, School of Engineering and Sciences, SRM University AP, Amaravati, 522240, India}
  \author{Rhonald Burgos Atencia}
  \email{rhburgos@gmail.com}
  \affiliation{Dirección de Ciencias Básicas, Universidad Tecnológica de Bolívar, Cartagena, Colombia}
  \author{Pankaj Bhalla}
  \email{pankaj.b@srmap.edu.in}
  \affiliation{Department of Physics, School of Engineering and Sciences, SRM University AP, Amaravati, 522240, India}
  \affiliation{Asian Consortium on Computational Materials Science—Global Research Centre, SRM University, Andhra Pradesh, AP 522240, India}
  
\date{\today}

\begin{abstract}
Our study investigates the orbital Hall (OH) conductivity originating from the disorder induced mechanisms, namely side-jump and skew scattering. On treating the disorder potential within the first-order Born approximation and beyond and employing the quantum kinetic approach, the general scaling law for both mechanisms in terms of disorder potential is provided to signal out the process experimentally. The calculations have been processed for the three-dimensional Weyl semimetals under an oscillating electric field. We reveal that the higher order disorder contributions leading skew scattering process dominates the orbital Hall conductivity over the intrinsic and side-jump processes.  In addition, we demonstrate the tunable nature of the extrinsic orbital Hall conductivity with the strength of the disorder potential and the external field. Finally, our findings highlights disorder scattering as a potential route for orbital transport.
\end{abstract}

\maketitle

\section{Introduction}
In recent years, the orbital degree of freedom of electrons instead of their spin and charge has emerged as a promising fundamental channel to generate and manipulate the orbital angular momentum (OAM) in materials ~\cite{Burgos_AP2024, Go2021, Yao_AOP2011, Padgett_OE2017, kittel_book2018, Bertlmann_book2023}, leading to the field of orbitronics. The important phenomenon in this direction is the generation of the orbital Hall effect (OHE) due to the transverse flow of electrons in the application of an external electric field. Moreover, the phenomenon is intrinsically tied with the orbital magnetic moment (OMM) by the self rotation of electron wave packets~\cite{Chang_JP2008, Sundaram_PRB1999, XiaoDi2010, Rhonald_PRB2026}. Interestingly, the OHE arises even in systems lacking a spin-orbit coupling, highlighting the origin of the effect from the orbital nature of the electronic states~\cite{Seifert_NN2023, Choi_N2023, Wang_NJP2023, Xue_PRB2020, Jo_PRB2018, Lee_CP2021, Zheng_PRR2020,  Kontani_PRL2008, Go_PRL2018, Salemi_PRM2022, Liu_PRL2025}. In addition, unlike the conventional charge Hall effect, which requires the breaking of space inversion and time reversal symmetry, the OHE can be notable in systems having preserved symmetries or broken either one of the symmetries, distinguishing the symmetry character of OHE~\cite{Kontani_PRL2008, Go_PRL2018, Busch_PRR2023, Sahu_PRB2021, Bhowal_PRB2021, Joy_PRB2025}. The sizable OH response is observed in a wide range of materials, which boosts the future orbitronic based technologies~\cite{Go2021, Bernevig_PRL2005}.

Basically, the OH response is led by two broad channels, such as field-driven and scattering. The intrinsic contribution to the system occurs by the interplay of band geometry and topological features of the electronic structure ~\cite{Lee_PRB2025, Bhowal_PRB2020, Bhowal_PRB2020M, Mankovsky_PRB2024}. The related intrinsic contribution to the OH response in both DC and AC regimes has been explored in systems such as Dirac and Weyl materials~\cite{Bhowal_PRB2021, Nagaosa_RMP2010, Bansil_RMP2016, Joy_PRB2025}. However, such contribution solely cannot provide the full description of orbital transport in real systems, where the scattering mechanism plays a significant role~\cite{Onoda_PRB2008, Culcer_PRB2017, Liu_PRL2024, Fert_PRL2011, Pandey_NJP2025, Ostrovsky_PRB2008, Shiomi_PRB2010, Chazalviel_PRB1974}. Yet the first experimental measurements on titanium thin films by Choi $\textit{et al.}$ argue that the overall OH response is affected by the scattering mechanisms~\cite{Choi_N2023, Veneri_PRL2025}. Peng \textit{et al.} show the dominant contribution of extrinsic contribution over the intrinsic to the overall OH response for strontium ruthenate system~\cite{Peng_NM2025}. In particular, the extrinsic contribution arises via two paths namely side jump and skew scattering. The side-jump scattering originates from the transverse coordinate shift of the electronic wave packet during impurity scattering and provides an anomalous displacement in the incident and final momentum~\cite{Burgos_AP2024, Liu_PRL2024, Peng_NM2025, Berger_PRB1970, Glazov_PRL2020, Yang_PRB2011, Shashank_PRB2023, Ma_PRB2025, Sinitsyn_PRB2005, Sinitsyn_PRB2006, Xiao_PRB2019}. In contrast, the skew scattering mechanism arises from asymmetric impurity scattering probabilities~\cite{Glazov_PRL2020, Ma_PRB2025, Ferreira_PRL2014, Wang_PRB2021, Guo_PRL2009, Herschbach_PRB2013, Niimi_PRL2012}. Recent theoretical predictions also reveal that the extrinsic contributions are essential and sometimes give dominant contribution to the OH response. Furthermore, on treating the disorder perturbatively within the first-order Born approximation, the extrinsic contribution does not show explicit disorder dependence to the OH response in all frequency regimes~\cite{Culcer_PRB2017, Liu_PRL2024, Ado_2015}. Moving beyond the first-order Born approximation where the higher order scattering processes take into account generates the disorder dependence of the OH response~\cite{Liu_PRL2024}. However, the progresses are restricted particularly to two-dimensional systems. This raises several important questions to have deeper insights of the role of extrinsic contributions to the OH response. How the disorder effects such as side-jump and skew scattering boost and manipulate the overall OH response in a three-dimensional system? Which extrinsic mechanism contributes more to the OH response? Can we expect some changes to the OH response by taking the impurity scattering to the higher order in the disorder potential analogues to the two-dimensional case ? To answer these questions, the present work extends the investigation of the OH response particularly by the scattering mechanisms in a three-dimensional system using a quantum kinetic approach. This provides a transparent framework for investigating the disorder-induced transport through the density matrix formulation. While the Keldysh Green's function formalism describes disorder effects in terms of self-energies and vertex corrections, the quantum kinetic approach is adopted here to facilitate the identification and physical interpretation of the distinct scattering contributions.
 
\begin{figure}[t]
      \includegraphics[width=1.0\linewidth]{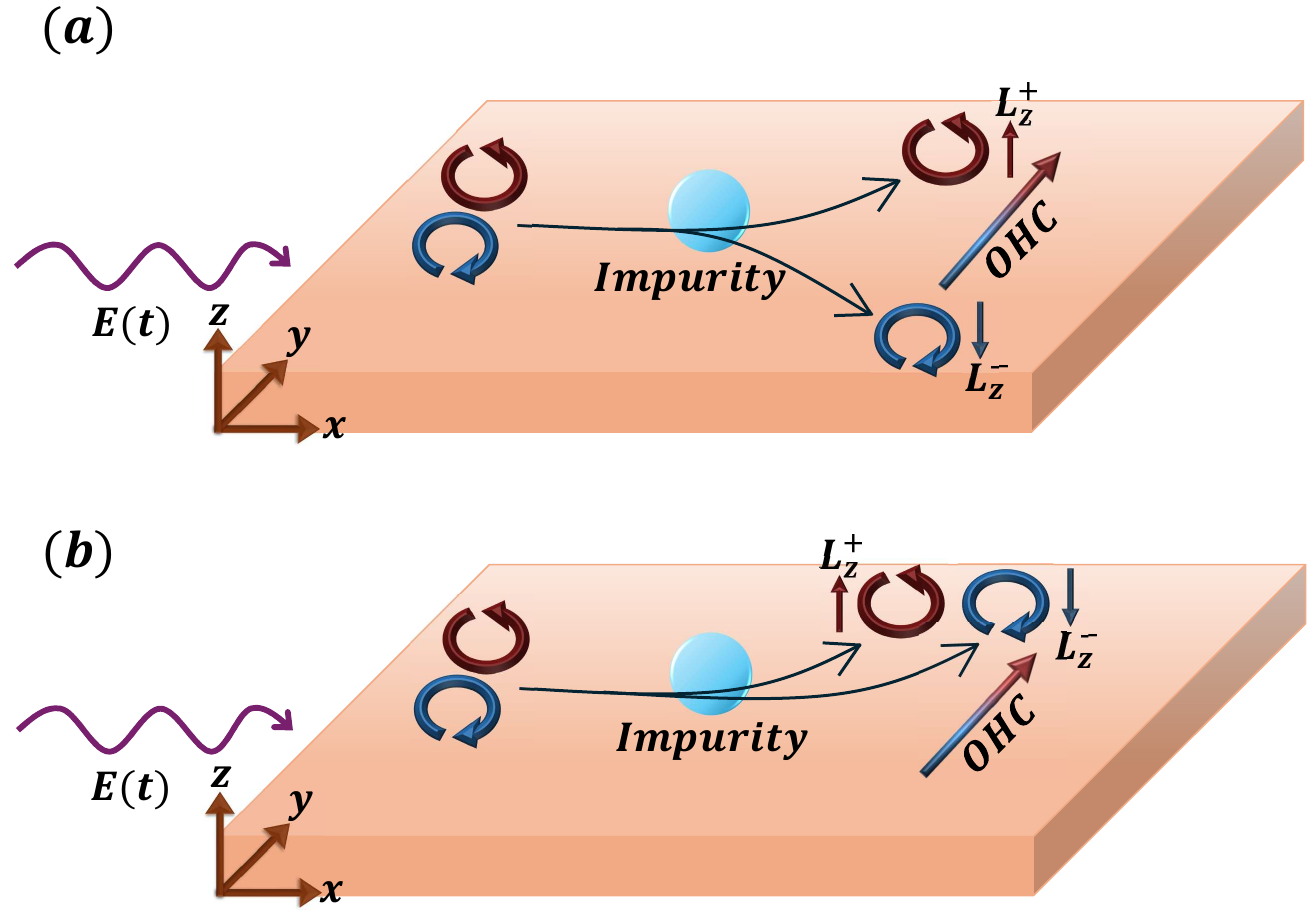}
    \caption{The diagram shows the orbital Hall current (OHC) generated due to the transverse motion of the orbital angular momentum (OAM) by the scattering process, such as skew scattering and side-jump mechanisms, under the influence of an oscillating electric field along the $x$-direction. (a) OHC induced by skew scattering mechanism: the impurity scatters the OAM ($L_z$) in asymmetrically forms, an accumulation of $L_z^{+}$ and $L_z^-$ in opposite directions, eventually generates OHC. (b) OHC induced from the side-jump mechanism: The impurity creates a coordinate shift in the OAM on the same side and generates OHC in the system.}
    \label{fig 2}
\end{figure}

Building on this framework, we systematically explore the two extrinsic mechanisms, side-jump and skew scattering, of OH conductivity, which are schematically represented in Fig.~\ref{fig 2}. For an applied electric field along the $x$-direction, asymmetric impurity scattering induces a transverse separation of OAM ($L_z$), leading to the orbital Hall current (OHC) associated with skew scattering as shown in Fig.~\ref{fig 2}(a). In contrast, the OHC from side-jump contribution due to the transverse shifts of the coordinates during the scattering events is portrayed in Fig.~\ref{fig 2}(b). Motivated by these considerations, the present work extends the investigation of AC-driven extrinsic OH conductivity in a three-dimensional Weyl semimetal system with broken time reversal symmetry using a quantum kinetic approach. Here, we demonstrate that the skew scattering dominates over the side-jump mechanism ($4-5$ orders of magnitude) in the extrinsic contribution to OH conductivity, with a disorder dependence of the $U_0^3$, and the extrinsic response is 10 times greater than the intrinsic part, in the lower regime of Weyl node separation and higher applied energy. The OH conductivity can be tuned by means of the strength of disorder potential ($U_0$), applied energy ($\hbar\omega$), and Weyl node separation ($b$). 

\section{Theoretical framework}
To compute OH conductivity, we employ the quantum kinetic equation framework, starting with the time-dependent quantum Liouville equation~\cite{Culcer_PRB2017, Bhalla_PRB2023, Pandey_PRB2024} for the density matrix $\rho$. Here the time evolution of $\rho$ reads as 
\begin{equation}
  \frac{\partial \rho}{\partial t} + \frac{i}{\hbar} [\mathcal{H},\rho]= 0,
  \label{QLE}
\end{equation} 
where $\mathcal{H}$ is the total Hamiltonian of the system and $\hbar$ is the reduced Planck's constant. Further, $\mathcal{H}$ is written as $\mathcal{H}(t)=\mathcal{H}_{\bm{q}}+\mathcal{H}_E(t)+U$, where $\mathcal{H}_{\bm{q}}$ is the band Hamiltonian, $\mathcal{H}_E(t)$ is the perturbed Hamiltonian due to an external electric potential, and $U$ stands for the disorder~\cite{Culcer_PRB2017, Liu_PRL2024, Bhalla_PRB2023}. For a spatially homogeneous and time-dependent electric field 
$\bm{E}(t)=E_0 e^{-i\omega t}$, the perturbation takes the form $\mathcal{H}_E(t)=e \bm{E}(t)\cdot \hat{\bm{r}}$, where $e$ is the charge of an electron, $\omega$ is the frequency and $\hat{\bm{r}}$ is the position vector of an electron. 

Here, we consider disorder potential for uncorrelated and randomly distributed impurity such that $U(\bm {r}) = \sum_{i} \delta(\bm{r} - \bm{r}_i)$ where $\bm{r}_i$ labels $i^{th}$ impurity location. 
For the disorder model we assume $\langle U(\bm{r})\rangle=0$ and $\langle U(\bm{r}) U(\bm{r'})\rangle=n_i\, U_0^2/\mathcal{V}\,\delta(\bm{r}-\bm{r'})$, where $U_0$ is the disorder strength having dimensions of energy times volume, $n_i$ refers to the impurity density, and $\mathcal{V}$ refers to volume which we consider as a unit throughout the calculation~\cite{Culcer_PRB2017, Veneri_PRL2025, Ma_PRB2025}.

The procedure to solve the kinetic equation is well described in Refs.~\cite{Culcer_PRB2017} and ~\cite{Atencia_arxiv2021}. It begins by decomposing the density matrix $\rho= \langle \rho \rangle + g$, where $\langle \rho \rangle$
is the disorder-averaged density matrix, and $g$ is a fluctuating part. We substitute this decomposition into the quantum Liouville equation Eq.~\eqref{QLE} and expand $\langle \rho \rangle = f_0 + f_E$, where $f_0$ is an equilibrium distribution function and $f_E$ is the field corrected part of the distribution function. Keeping the kinetic equation within the linear order in the electric field and expressing in the crystal momentum representation, one obtains the differential equation for $f_{E,{\bm q}}$ as 
\begin{equation}
\frac{\partial f_{E,{\bm q}}}{\partial t} + \frac{i}{\hbar} [\mathcal{H}_{\bm q},f_{E,{\bm q}}] + \frac{i}{\hbar} [\mathcal{H}_{E},f_{0,{\bm q}}] +\mathcal{J}_{\bm q}= 0.
\label{Eq.4}
\end{equation}
Here $\mathcal{J}_{\bm q}$ is the collision integral defined within the Born approximation as 
%
\begin{equation}
    \mathcal{J}_{\bm q} = \frac{1}{\hbar^2} \int_{-\infty}^{t} dt_1 \langle \big[U,  \big[e^{-i\mathcal{H}_{\bm q} t_1/\hbar} U e^{i\mathcal{H}_{\bm q} t_1/\hbar} , f_{E,{\bm q}} \big]\big] \rangle. 
    \label{Eq.J}
\end{equation}
Note that here $f_{E,{\bm q}}$ is also a function of time which comes from the field ${\bm E}(t)$. For the band dynamics, we will proceed now onward within the band basis representation. With this, the commutation relation of $f_{E,{\bm q}}$ and band Hamiltonian yields $\langle m|[\mathcal{H}_{\bm q},f_{E,{\bm q}}]|n\rangle=(\varepsilon_{\bm{q}}^{m}-\varepsilon_{\bm{q}}^{n})f_{E,{\bm q}}^{mn}$, reflecting the energy difference between the bands $m$ and $n$ where $|m\rangle = e^{-i{\bm q}\cdot {\bm r}}|u_{\bm q}^{m}\rangle$ with $| u_{\bm q}^{m}\rangle$ as the periodic part of the Bloch wave function. Here, $f_{E,{\bm q}}^{mn} = \langle u_{\bm q}^{m} | f_{E,{\bm q}} | u_{\bm q}^{n} \rangle$ and $\varepsilon_{\bm q}^{m}$ represents the band energy associated with the band index $m$. Clearly, for intraband case $m=n$, the relation $\langle m|[\mathcal{H}_{\bm q},f_{E,{\bm q}}]|n\rangle=0$. 

The driving term $\langle m|[\mathcal{H}_{\bm{E}},f_0]|n\rangle$ yields $ie\bm{E}(t)\partial_{\bm{q}}f_{0}^m \delta_{mn} + e \bm{E}(t)\mathcal{R}_{\bm{q}}^{mn}(f_{0}^{n}-f_{0}^{m})$, where $\mathcal{R}_{\bm{q}}^{mn}=i\langle u_{\bm q}^{m}|\partial_{\bm{q}}u_{\bm q}^{n}\rangle$ is the Berry connection of the system, $\partial_{\bm{q}}$ is the partial derivative with respect to momentum ${\bm q}$. To obtain it, we use $\hat{{\bm r}}|m\rangle=[i\partial_{\bm q} e^{-i{\bm q}\cdot {\bm r}}]|u_{\bm q}^{m}\rangle$. Further, $f_{0,{\bm q}}^{m}=[e^{(\varepsilon_{\bm {q}}^{m} - \varepsilon_f)/k_B T} +1]^{-1}$ is the equilibrium Fermi-Dirac distribution function associated with energy corresponding to band index $m$, $T$ is the absolute temperature, $k_B$ is the Boltzmann constant and $\varepsilon_f$ as the Fermi energy. The first term of the driving part here accounts for the intraband dynamics, and the second term governs the interband $m \neq n$ dynamics and includes the quantum geometrical picture via the Berry connection. To evaluate the collision integral explicitly, we transform Eq.~\eqref{Eq.J} in the eigenstates basis. After integration over time~\cite{Ma_PRB2025}
\begin{widetext}
\begin{equation}
\mathcal{J}^{mp}_{\bm{q}}=\frac{i}{\hbar}\sum_{n, l}\sum_{{\bm q}'}\bigg\{\frac{\langle U_{\bm{qq'}}^{mn}U_{\bm{q'q}}^{lp}\rangle f_{E,\bm{q'}}^{nl}}{\varepsilon_{\bm{q'}}^{l}-\varepsilon_{\bm{q}}^{p}-i\eta}-\frac{\langle U_{\bm{qq'}}^{mn}U_{\bm{q'q}}^{nl}\rangle f_{E,\bm{q}}^{lp}}{\varepsilon_{\bm{q'}}^{n}-\varepsilon_{\bm{q}}^{l}-i\eta}+\frac{\langle U_{\bm{qq'}}^{mn}U_{\bm{q'q}}^{lp}\rangle f_{E,\bm{q'}}^{nl}}{\varepsilon_{\bm{q}}^{m}-\varepsilon_{\bm{q'}}^{n}-i\eta}-\frac{f_{E,\bm{q}}^{mn}\langle U_{\bm{qq'}}^{nl}U_{\bm{q'q}}^{lp}\rangle}{\varepsilon_{\bm{q}}^{n}-\varepsilon_{\bm{q'}}^{l}-i\eta}\bigg\}.
\label{CT1}
\end{equation}
\end{widetext}
Here, $\langle U_{\bm q\bm q'}^{mn}\rangle=\langle u_{\bm q}^m|U|u_{\bm q'}^n\rangle$ denotes the disorder averaged matrix element of disorder potential. It is evident from the above equation that the scattering term is proportional to the second-order potential, i.e., $\mathcal{J}\propto U_0^2$. Further, the factor $\eta$ is infinitesimally small and comes while regularising the time integral. The details of the average of product of disorder potentials are given in Appendix~\ref{App:C}.

Based on Eq.~\eqref{CT1}, the collision integral depends on the field-induced components of the density matrix. This leads to the collision integral into two different scenarios. First, the off-diagonal part of the collision integral gives $\mathcal{J}_{\bm{q}}^{mp}[f_{\bm E,{\bm q}}] = \mathcal{J}_{\bm{q}}^{mp}[f_{\bm E,{\bm q}}^{mm}]+\mathcal{J}_{\bm{q}}^{mp}[f_{\bm E,{\bm q}}^{mp}]$ where $\mathcal{J}_{\bm{q}}^{mp}[f_{\bm E,{\bm q}}^{mm}]$ part incorporates the scattering contributions through the intraband components of the density matrix and this leads to side-jump contribution as discussed below. The other part has been treated within the relaxation time approximation as $f_{E,{\bm q}}^{mp}/\tau$ having $\tau$ a constant. Second, the diagonal part of the collision integral yields $\mathcal{J}_{\bm{q}}^{mm}[f_{\bm E,{\bm q}}] = \mathcal{J}_{\bm{q}}^{mm}[f_{\bm E,{\bm q}}^{mm}]+\mathcal{J}_{\bm{q}}^{mm}[f_{\bm E,{\bm q}}^{mp}]$. Here, the term $\mathcal{J}_{\bm{q}}^{mm}[f_{\bm E,{\bm q}}^{mp}]$ gives the another channel for the extrinsic contribution that can lead to anomalous Hall effect~\cite{Ado2017-2} and the other term is considered as $f_{\bm E,{\bm q}}^{mm}/\tau_{\bm q}^{m}$. 
The detailed expressions for the scattering term are provided in Appendix~\ref{App:A}. Substituting the expression for the scattering term back into Eq.~\eqref{Eq.4} and keeping $\mathcal{J}$ within the first-order Born approximation, the linear-order differential equation for the off-diagonal components of the side-jump correction of the density matrix can be solved. 
The procedure is as follows. Eq.~\eqref{Eq.4} is solved in perturbation theory in both the electric field and impurity scattering. First for the diagonal part, the leading distribution follows from ~\cite{Culcer_PRB2017}
\begin{equation}
\mathcal{J}_{\bm q}^{mm}=-\frac{i}{\hbar} [\mathcal{H}_{E},f_{0,{\bm q}}]^{mm}
\label{eq:5}
\end{equation}
which leads to the Boltzmann distribution that we call $f_{E,{\bm q}}^{mm}$. This distribution does not contribute to the OH response but is the starting point to calculate the sub-leading distributions leading to the OHE. The sub-leading distribution has both diagonal and off-diagonal contributions. The off-diagonal contributions follow from the kinetic equation 
\begin{equation}
\frac{\partial f^{mp}_{E,{\bm q}}}{\partial t} + \frac{i}{\hbar} [\mathcal{H}_{\bm q},f_{E,{\bm q}}]^{mp} = - \frac{i}{\hbar} [\mathcal{H}_{E},f_{0,{\bm q}}]^{mp} - \mathcal{J}^{mp}(f_{E,{\bm q}}).
\label{Eq:keq}
\end{equation}

The first driving term on the right hand side gives rise to a purely intrinsic contribution, namely 
\begin{equation}
f_{E,{\bm q}}^{\mathrm{int}, mp} = - \frac{ \frac{i}{\hbar} [\mathcal{H}_{E},f_0]^{mp} }{ (\frac{i}{\hbar}(\varepsilon_{\bm{q}}^{m}-\varepsilon_{\bm{q}}^{p})-i\omega)},
\end{equation}
while the second driving term gives rise to a side jump effect \cite{Sinitsyn_2007JPCM, Sinitsyn_2007}, which is the off-diagonal component of the collision integral $\mathcal{J}^{mp}(f_{E,{\bm q}}^{mm})$, namely,
\begin{equation}
f_{E,{\bm q}}^{\mathrm{sj}, mp} = - \frac{ \mathcal{J}^{mp}(f_{E,{\bm q}}^{mm})}{(\frac{i}{\hbar}(\varepsilon_{\bm{q}}^{m}-\varepsilon_{\bm{q}}^{p})-i\omega)}.
\end{equation}

Clearly, the collision integral $\mathcal{J}^{mp}(f_{E,{\bm q}}^{mm})$ is proportional to the product of $U_0^2$ and $f_{E,\bm{q}}^{mm}$. Further, the distribution function is 
$f_{E,{\bm q}}^{mm}=e\bm{E}(t)\partial_{\bm{q}}f_{0}^m/(\omega + i/\tau_{\bm q}^m)$. Here, $\tau_{\bm q}^{m}$ is the transport time scale for short range impurities and expressed as 
\begin{equation}
\frac{1}{\tau_{\bm q}^{m}} =\frac{2\pi}{\hbar} \sum_{{\bm q}'}\sum_{p} \delta(\varepsilon_{\bm q}^m - \varepsilon_{{\bm q}'}^p) \langle U_{\bm {qq}'}^{mp}U_{\bm{q}'\bm q}^{pm}\rangle (1-\cos(\phi_{\bm q}-\phi_{\bm q'})), 
\end{equation}
where $\phi_{\bm q}$ denotes the angle obtained by momentum vector $\bm q$ along $x-$direction, while $(\phi_{\bm q}-\phi_{{\bm q}'})$ indicates the scattering angle between two states~\cite{Culcer_PRB2017, Atencia_arxiv2021}. Clearly, $\tau_{\bm q}^{m}$ here varies as $\propto U_0^{-2}$. 
Depending on the different frequency regimes, the scaling behavior of the collision integral with the disorder potential strength is discussed below.

In the low-frequency regime, $\omega\tau\ll1$, in which the diagonal non-equilibrium  distribution function is proportional to transport time contribution, i.e., $f_{E,{\bm q}}^{mm}\propto\tau_{\bm q}^{m}\propto U_0^{-2}$. From the above Eq.~\eqref{CT1}, the off-diagonal collision integral exhibits an explicit disorder dependence through the disorder average $\langle U_{\bm{qq'}}^{mn}U_{\bm{q'q}}^{np}\rangle\propto U_0^2$ and an implicit disorder dependence through the diagonal non-equilibrium  distribution function. Therefore, the potential dependence in the collision integral nullifies and consequently $f_{E,{\bm q}}^{\mathrm{sj}, mp}$ gives potential independent behavior, as $f_{E,{\bm q}}^{\mathrm{sj}, mp}\propto\mathcal{J}^{mp}_{\bm{q}}(f_{E,{\bm q}}^{mm})\propto U_0^0$. 

Besides the side jump effect, there is one more contribution which is extrinsic and disorder independent. It is given by the equation 
\begin{equation}
f_{E,{\bm q}}^{\mathrm{IC},mm} = - \frac{\mathcal{J}^{mm}(f_{E, {\bm q}}^{mp})}{(1/\tau_{\bm q}^m + i\omega)}.
\end{equation}
Where $f_{E,{\bm q}}^{\mathrm{IC},mm}$ is the interband coherence induced density matrix within the first Born approximation. Here, the $f_{E, {\bm q}}^{mp}$ is independent of disorder strength and is written in the form $ e \bm{E}\cdot\mathcal{R}_{\bm{q}}^{mp}(f_{0}^{m}-f_{0}^{p})/(\varepsilon_{\bm q}^m-\varepsilon_{\bm q}^p-\hbar\omega)$. The disorder strength dependence on the $f_{E,{\bm q}}^{\mathrm{IC},mm}$ is explicitly from the collision integral ($\mathcal{J}^{mm}$) and the transport time $1/\tau_{\bm q}^m$, i.e., $f_{E,{\bm q}}^{\mathrm{IC},mm}\propto\mathcal{J}^{mm}\tau_{\bm q}^m \propto U_0^0$. This contribution is in correspondence with the scattering contribution reported in the anomalous Hall effect \cite{Ado2017-2}.

Moreover, at high frequencies, where the interband transitions are valid, the weak disorder condition $\hbar/\tau\ll\varepsilon_f$ ensures that the disorder broadening remains small. In this limit, where $1/\tau_{\bm q}^m + i\omega\approx i\omega$, the density matrix corrections arising from the Born approximation collision integral scale as $U_0^2$, reflecting the quadratic dependence of the disorder potential strength.

\subsection{Beyond the first Born approximation}
The intrinsic and side jump contributions are independent of disorder strength and are captured within the first Born approximation. However, some experiment related to the anomalous Hall effect show that there is an explicit dependence with disorder strength \cite{TianYann2009}, not capture with the theoretical description so far explained. 
This suggest a possible similar behavior in the orbital Hall effect.
To find this leading-order dependence of the disorder potential on response, one has to go beyond the first-order Born approximation \cite{Sinitsyn_2007}. For this, we
extend the scattering term $\mathcal{J}$ to the third-order $U$ which is expressed as
\begin{equation}\begin{aligned}
\mathcal{J}_{1,{\bm q}} =-\frac{i}{\hbar^3}\int_{-\infty}^{t}dt_1\int_{-\infty}^{t}dt_2&\langle[U,[e^{-i\mathcal{H}_{\bm q} t_1/\hbar}Ue^{i\mathcal{H}_{\bm q} t_1/\hbar},\\&[e^{-i\mathcal{H}_{\bm q} t_2/\hbar}Ue^{i\mathcal{H}_{\bm q} t_2/\hbar},f_E]]]\rangle.\end{aligned}
\end{equation}
To evaluate the third-order collision integral, we consider the disorder average of the product of three impurity potential matrix elements. The average can by written as $\langle U_{\bm q\bm q'}^{mm'}U_{\bm q'\bm q''}^{m'm''}U_{\bm q''\bm q'''}^{m''m'''}\rangle$. 
Keeping the contribution in which all three scattering events occur at the same impurity. To distinguish the above equation from the first-order Born approximated $\mathcal{J}$, we use the notation for the scattering term here as $\mathcal{J}_{1,{\bm q}}$. 
Within the band basis representation, 
$\mathcal{J}_{1,{\bm q}}^{mp}$~\cite{Yang_PRB2011, Ma_PRB2025} is given in the form
\begin{widetext}
\begin{equation}
\begin{aligned}
    \mathcal{J}_{1,\bm{q}}^{mp}=
    \frac{i}{\hbar}\sum\limits_{n,l,k}\sum\limits_{\bm q' \bm q''}&\bigg\{\frac{\langle U_{\bm{qq'}}^{mn}U_{\bm{q'q''}}^{nl}U_{\bm{q''q}}^{lk}\rangle f_{E,\bm{q}}^{kp}}{(\varepsilon_{\bm{q'}}^{n}-\varepsilon_{\bm{q''}}^{l}-i\eta)(\varepsilon_{\bm{q''}}^{l}-\varepsilon_{\bm{q}}^{k}-i \eta)}- \frac{f_{E,\bm{q}}^{mn}\langle U_{\bm{qq'}}^{nl}U_{\bm{q'q''}}^{lk}U_{\bm{q''q}}^{kp}\rangle}{(\varepsilon_{\bm{q'}}^{l}-\varepsilon_{\bm{q''}}^{k}-i\eta)(\varepsilon_{\bm{q}}^{n}-\varepsilon_{\bm{q'}}^{l}-i\eta)}\\&- \frac{\langle U_{\bm{qq'}}^{mn}U_{\bm{q'q''}}^{nl}U_{\bm{q''q}}^{kp}\rangle f_{E,\bm{q''}}^{lk}}{(\varepsilon_{\bm{q'}}^{n}-\varepsilon_{\bm{q''}}^{l}-i\eta)(\varepsilon_{\bm{q''}}^{k}-\varepsilon_{\bm{q}}^{p}-i\eta)}-\frac{\langle U_{\bm{qq'}}^{mn}U_{\bm{q'q''}}^{nl}U_{\bm{q''q}}^{kp}\rangle f_{E,\bm{q''}}^{lk}}{(\varepsilon_{\bm{q''}}^{k}-\varepsilon_{\bm{q}}^{p}-i\eta)(\varepsilon_{\bm{q'}}^{n}-\varepsilon_{\bm{q''}}^{l}-i\eta)}\\&-\frac{\langle U_{\bm{qq'}}^{mn}U_{\bm{q'q''}}^{nl}U_{\bm{q''q}}^{kp}\rangle f_{E,\bm{q''}}^{lk}}{(\varepsilon_{\bm{q}}^{m}-\varepsilon_{\bm{q'}}^{n}-i\eta)(\varepsilon_{\bm{q'}}^{n}-\varepsilon_{\bm{q''}}^{l}-i\eta)}+\frac{\langle U_{\bm{qq'}}^{mn}U_{\bm{q'q''}}^{lk}U_{\bm{q''q}}^{kp}\rangle f_{E,\bm{q'}}^{nl} }{(\varepsilon_{\bm{q''}}^{k}-\varepsilon_{\bm{q}}^{p}-i\eta)(\varepsilon_{\bm{q'}}^{l}-\varepsilon_{\bm{q''}}^{k}-i\eta)}\\&+ \frac{\langle U_{\bm{qq'}}^{mn}U_{\bm{q'q''}}^{lk}U_{\bm{q''q}}^{kp}\rangle f_{E,\bm{q'}}^{nl}}{(\varepsilon_{\bm{q}}^{m}-\varepsilon_{\bm{q'}}^{n}-i\eta)(\varepsilon_{\bm{q'}}^{l}-\varepsilon_{\bm{q''}}^{k}-i\eta)}+ \frac{\langle U_{\bm{qq'}}^{mn}U_{\bm{q'q''}}^{lk}U_{\bm{q''q}}^{kp}\rangle f_{E,\bm{q'}}^{nl}}{(\varepsilon_{\bm{q'}}^{l}-\varepsilon_{\bm{q''}}^{k}-i\eta)(\varepsilon_{\bm{q}}^{m}-\varepsilon_{\bm{q'}}^{n}-i\eta)}\bigg\}.
    \end{aligned}
\end{equation}
\end{widetext}
The third-order contribution to the scattering term $ \mathcal{J}_{1,\bm{q}}^{mp}$ arises from the successive impurity scattering and is therefore proportional to $U_0^3$, as is evident from the above expression. The further details are given in Appendix~\ref{App:C}. The skew scattering correction to the density matrix, obtained by employing the third-order collision integral in the quantum kinetic equation, is denoted as $f_{1E, \bm q}^{\text{sk}}$. 
For the diagonal component, the correction term takes the form 
\begin{equation}
f_{1E, \bm q}^{\text{sk},mm}=-\frac{\mathcal{J}_1^{mm}(f_{E,{\bm q}}^{mp})}{1/\tau_{\bm q}^{m}+i\omega}.
\label{Eq:13}
\end{equation}
In the low frequency region, the denominator is dominated by the transport time, this leads to $f_{1E, \bm q}^{\text{sk},mm}\propto\mathcal{J}_1^{mm}(f_{E,{\bm q}}^{mp})\tau_{\bm q}^{m}\propto U_0^3U_0^{-2}$. Thus, skew scattering density matrix exhibits a linear dependence on the disorder potential, $f_{1E, \bm q}^{\text{sk},mm}\propto U_0$. Similarly, the off-diagonal correction term contains the diagonal intrinsic density matrix in the collision term, i.e., 
\begin{equation}
f_{1E,{\bm q}}^{\text{sk}, mp} = -\frac{\mathcal{J}_1^{mp}(f_{E,{\bm q}}^{mm})}{(\frac{i}{\hbar}(\varepsilon_{\bm{q}}^{m}-\varepsilon_{\bm{q}}^{p})-i\omega)}.
\label{Eq:14}
\end{equation} Here, the linear dependence of $f_{1E,{\bm q}}^{\text{sk}, mp}$ on $U_0$ expressed as $f_{1E,{\bm q}}^{\text{sk}, mp} \propto \mathcal{J}_1^{mp}(f_{E,{\bm q}}^{mm})\propto U_0$ again traces from the factor $\tau_{\bm q}^{m}$ inherited from the diagonal component of the intrinsic density matrix. 

On the other hand in the high frequency regime, the third-order collision integral gives $\mathcal{J}_{1}\propto U_0^3$. Consequently, the diagonal and off-diagonal skew scattering density matrix corrections $f_{1E, \bm q}^{\text{sk},mm}$ and $f_{1E,{\bm q}}^{\text{sk}, mp}$ gives a cubic dependence on the disorder potential. 

It is important to note that the extrinsic contribution here is restricted to the weak disorder limit, where the impurity potential is sufficiently small under the disorder broadening condition $\hbar/\tau \ll \varepsilon_f$. Therefore, the higher-order scattering contributions such as $U_0^4, U_0^5$ and beyond are neglected, as they are expected to be progressively suppressed in weak disorder. 
\subsection{Orbital Hall current}
OHE describes the generation of a transverse motion of OAM in response to an applied electric field. Here, we focus on impurity-induced contributions such as side-jump and skew scattering to the OHC. To evaluate, we employ the definition of the OHC, 
\begin{equation}
J_\alpha=\text{Tr}(\hat{j}_\alpha f_{{E,\bm{q}}}),
\end{equation}
where $\hat{j}_\alpha=\frac{1}{2}\{\hat{L}_\beta,\hat{v}_\alpha\}$ is the OHC operator in an arbitrary $\alpha$-direction~\cite{Bhowal_PRB2021, Joy_PRB2025, Liu_PRL2024, Joy_NJP2026}. Here, $\hat{L}_\beta$ denotes the OAM in the $\beta$-direction and $\hat{v}_\alpha$ is the velocity operator along the $\alpha$-direction. The OAM elements represents in the two-band basis, $\hat{\bm L}_\beta^{mp}=\langle u_{\bm q}^m|\hat{\bm L}_{\beta}|u_{\bm q}^p\rangle$, where the matrix representation depends on the orbital character of the basis states. In a general way, the OAM proportional to the vector product of the position and velocity operators is expressed as 
\begin{equation}
    \hat{L}_{\beta}=\frac{e\hbar}{4g_l\mu_B}(\hat{r}_{\gamma}\hat{v}_{\delta}-\hat{v}_{\gamma}\hat{r}_{\delta}),
\end{equation}
where, $\mu_B$ is the Bohr-magneton, $\alpha$, $\beta$, $\delta$ and $\gamma$ are arbitrary directions and $g_{l}$ is the Land\'{e} g-factor of OAM. The OHC contains the position and velocity operators. In the Bloch eigenstate basis, the interband matrix elements of position and velocity operators contain the Berry connection. 
\begin{equation}\begin{aligned}
    \hat{\bm j}_{\alpha}^{mp} =\frac{e\hbar}{8g_l\mu_B}\sum\limits_{m,n,l,p} \bigg(\hat{\bm r}_{\gamma}^{mn}\hat{\bm v}_{\delta}^{nl}\hat{\bm v}_{\alpha}^{lp}-\hat{\bm v}_{\gamma}^{mn}\hat{\bm r}_{\delta}^{nl}\hat{\bm v}_{\alpha}^{lp}\\+\hat{\bm v}_{\alpha}^{mn}\hat{\bm r}_{\gamma}^{nl}\hat{\bm v}_{\delta}^{lp}-\hat{\bm v}_{\alpha}^{mn}\hat{\bm v}_{\gamma}^{nl}\hat{\bm r}_{\delta}^{lp}\bigg).
\label{Jy}
\end{aligned}\end{equation}
Where the $\hat{\bm{r}}_{\bm{qq}'}^{mn}=  \sum\limits_{q'}\big(i\frac{\partial \delta(\bm{q}-\bm{q}')}{\partial \bm{q}'}\delta_{mn}+ \mathcal{R}_{\bm{q}'}^{mn}\delta(\bm{q}-\bm{q}')\big)$, and $\hat{\bm{v}}_{\bm{q}}^{mn}=\frac{1}{\hbar}\frac{\partial \varepsilon_{\bm{q}}^m}{\partial \bm{q}}\delta_{mn} + \frac{i}{\hbar}\ \mathcal{R}_{\bm{q}}^{mn}\ (\varepsilon_{\bm{q}}^{m}-\varepsilon_{\bm{q}}^{n})$. The products of interband Berry connections from the position and velocity operators can be written in terms of the quantum geometric tensor (QGT), defined as $\mathcal{Q}_{ab}^{mp}=\mathcal{R}^{mp}_a\mathcal{R}_{b}^{pm}$. Further, the field corrected density matrix part $f_{E,{\bm{q}}} = f_{E,{\bm{q}}}^{\text{sj}} + f_{E,{\bm{q}}}^{\text{sk}} + f_{1E,{\bm{q}}}^{\text{sk}}$ leads the orbital current in the form
\begin{equation}\begin{aligned}
    J_\alpha = \sum_{m,p} \sum_{\bm q} \hat{j}_{\alpha}^{mp} f_{E,{\bm{q}}}^{pm}.
    \end{aligned}\end{equation}
Writing OHC explicitly for different contributions, we get
\begin{equation}\begin{aligned}
    J_\alpha = \sum_{\bm q}\bigg\{ &\sum_{m} \big( \hat{j}_{\alpha}^{mm} f_{E,{\bm{q}}}^{\text{sk},mm}+\hat{j}_{\alpha}^{mm} f_{1E,{\bm{q}}}^{\text{sk},mm}\big)\\&+\sum_{p\neq m} \big(\hat{j}_{\alpha}^{mp} f_{1E,{\bm{q}}}^{\text{sk},pm}+\hat{j}_{\alpha}^{mp} f_{E,{\bm{q}}}^{\text{sj},pm} \big)\bigg\}.
    \end{aligned}\end{equation}
\begin{figure*}[t]
      \includegraphics[width=1.0\linewidth]{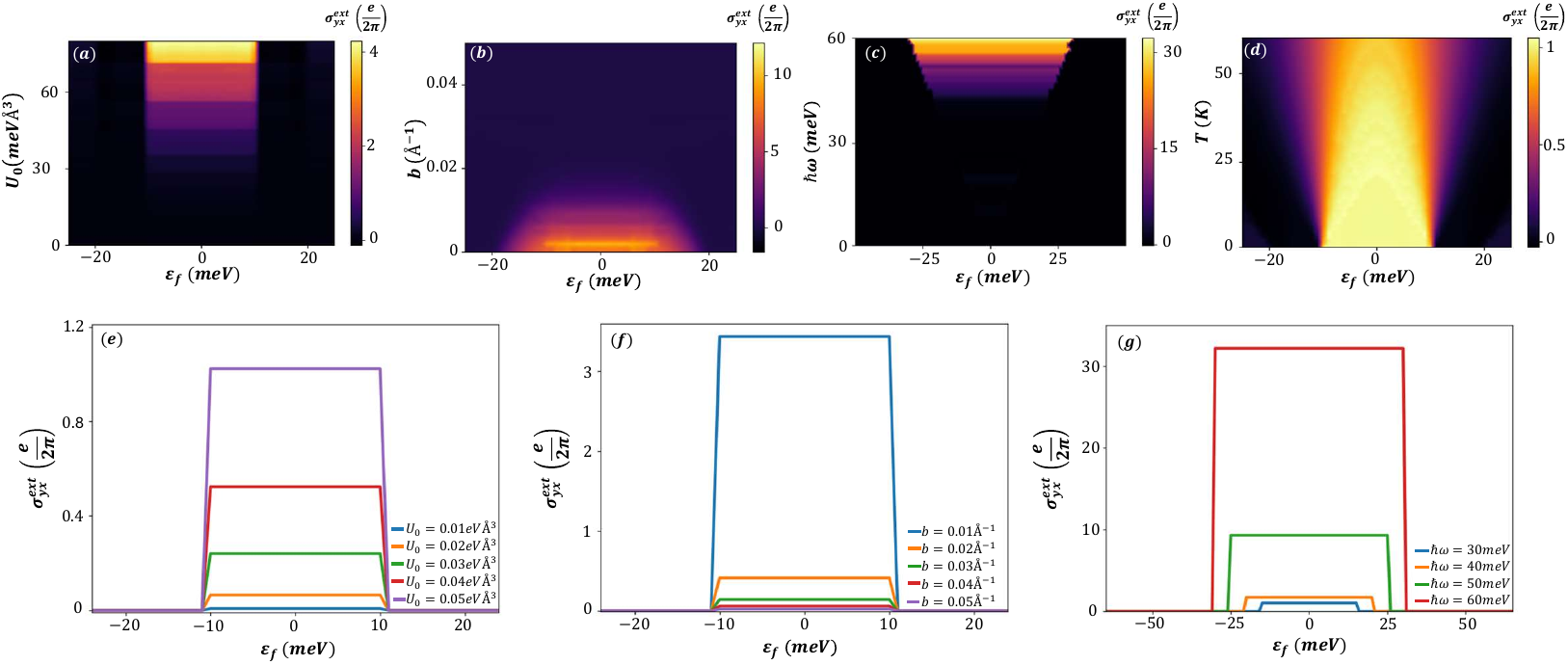}
    \caption{The figure illustrates the dependence of total OH conductivity $(\sigma_{yx}^{ext})$ on Fermi energy $(\varepsilon_f)$ for different parameters. Panels (a), (b), (c), and (d) show the variation of $\sigma_{yx}^{ext}$ with strength of disorder potential ($U_0$), Weyl node separation ($b$), applied energy ($\hbar\omega$), and the temperature, respectively. In each panel, the remaining parameters are fixed at $b=0.04\,\text{\AA}^{-1}$, $\hbar\omega=20\,\text{meV}$, $U_0=50\,\text{meV\AA}^3$ and $T=1$~K. In panels (e), (f), and (g), the variation of total extrinsic OH conductivity as a function of Fermi energy is depicted in two-dimensional plot.}
    \label{fig 1}
\end{figure*}
\section{Model and Discussion} 
To investigate the extrinsic contribution to the OH conductivity, we consider a three-dimensional time reversal symmetry broken Weyl semimetal described by an effective $\bm k\cdot \bm p$ two-band band Hamiltonian which is given below~\cite{Ahn_PRB2017, Yang_PRL2015}, 
 \begin{equation}
     \mathcal{H}_{\bm{q}}=\hbar v_f\zeta[q_x \sigma_x + q_y \sigma_y + (q_z - \zeta b)\sigma_z].
     \label{Eq.1}
 \end{equation}
Here, the orbital basis Pauli matrix is referred as $\sigma_{\alpha}$, and the Fermi velocity is $v_f$. The pair of Weyl nodes having opposite chiralities represented by $+1$ and $-1$ which is denoted by chiral index 
$\zeta$ in the Hamiltonian. The Weyl node separation is indicating the momentum space shift of nodes along the $z$-direction characterised by the term $b$~\cite{Zyuzin_PRB2012, Burkov_PRL2011}, which is within the first Brillouin zone consistent to the experimental values~\cite{Huang_NC2015, Grassano_SR2018, Xu_Science2015, Levy_PRB2020}. 
The energy spectrum cooresponding to the Hamiltonian is given by $\varepsilon_{\bm q}^{\pm}=\pm v_f\hbar\zeta\sqrt{q_x^2+q_y^2+(q_z-\zeta b)^2}$. The eigenstates required to evaluate the Berry connection and the collision integral take the form,
\begin{equation}
|u_{\bm q}^{\pm}\rangle
=
\frac{1}{\sqrt{2}}
\begin{pmatrix}
\sqrt{1\pm\frac{d_z}{d}}
\\
\pm
\frac{(q_x+i q_y)}
{d
\sqrt{1\pm\frac{d_z}{d}}}
\end{pmatrix},
\end{equation}
where $d_z=(q_z-\zeta b)$ and $d=\sqrt{q_x^2+q_y^2+d_z^2}$. The two basis states represent the internal degrees of freedom incorporated in the effective low energy model. The OHC operator corresponding to the current flows along the $y$-direction with OAM in $z$-direction $\hat{j}_y=\frac{1}{2}\{\hat{L}_z,\hat{v}_y\}$, can be represented in two band basis as, 
$\hat{j}_y^{\pm\pm}=\frac{1}{2}(\hat{L}_z^{\pm\pm}\hat{v}_y^{\pm\pm} +\hat{L}_z^{\pm\mp}\hat{v}_y^{\mp\pm})$ and $\hat{j}_y^{\pm\mp}=\frac{1}{2}(\hat{L}_z^{\pm\pm}\hat{v}_y^{\pm\mp} +\hat{L}_z^{\pm\mp}\hat{v}_y^{\mp\mp})$. The corresponding matrix element of the OHC operator on substituting the position and velocity matrix elements for the considered Hamiltonian~\eqref{Eq.1} within two-band basis is expressed as
\begin{equation}
\begin{aligned}
  &\hat{j}_{y}^{\pm\pm}=\frac{v_f^2 (q_y+i q_x)}{2 q_\perp^2}\pm\frac{v_f^2 q_yd_z}{2d^3},\,\hat{j}_{y}^{\pm\mp}=\frac{-v_f^2 q_y d_z^2}{2 d^3 q_\perp}\pm\frac{i v_f^2 q_x d_z}{2 d^2 q_\perp},
  \end{aligned}
  \end{equation} respectively. Here, $q_\perp=\sqrt{q_x^2+q_y^2}$ denotes the momentum component transverse to the $q_z$-axis. 

From our evaluation for time reversal broken Weyl semimetal, we find the total extrinsic OH conductivity is dominated by the skew scattering mechanism stemming from beyond first-order Born approximation, thus $\sigma_{yx}^{ext}\approx\sigma_{yx}^{sk}$. Note that the skew scattering contribution within the first Born approximation is independent of disorder strength and provides a negligible contribution to the total extrinsic OH conductivity. Furthermore, the skew scattering induced total extrinsic OH conductivity ($\sigma_{yx}^{ext}$) shows a linear dependence with the disorder potential, as shown in the density plot for $\sigma_{yx}^{ext}$  with $U_0$ and $\varepsilon_f$ in Fig.~\ref{fig 1}(a) and in two-dimensional plot Fig.~\ref{fig 1}(e). Here, the response is displayed in terms of $e/2\pi$ factor. This linear behavior to the OHC results from the skew scattering contribution stemming from field correction part $f_{1E}^{\text{sk}}$ Eqs.~\eqref{Eq:13} and ~\eqref{Eq:14}. Notably, $\sigma_{yx}^{ext}$ displays a plateau feature around the $\hbar\omega=|2\varepsilon_f|$. Here, the plateau is visible around $\varepsilon_f=\pm10\,\text{meV}$ for $\hbar\omega=20 \text{meV}$. This behavior originates from the threshold condition of interband transitions, which occurs when $\hbar\omega=|2\varepsilon_{f}|$. In this regime, the OH response is constant over the Fermi energy, giving rise to the observed stable segment. 

Fig.~\ref{fig 1}(b) and~(f) illustrates the extrinsic OH conductivity for different Weyl node separation $(2b)$. Here, we have taken $b$ within the first Brillouin zone. The results indicate that the extrinsic part of the OH conductivity suppresses with the node separation, in contrast to the intrinsic contribution~\cite{Joy_NJP2026}. On further reducing the Weyl node separation, the system moves from the Weyl to the Dirac semimetal limit, i.e., with $b$ approaches zero, and the two Weyl nodes merge into one. Here, the intrinsic contribution, which is associated with band geometry, decreases with decreasing $b$ and vanishes at the Dirac limit, where both time and inversion symmetry are preserved~\cite{Joy_PRB2025}. In contrast, the extrinsic contribution to the OH conductivity remains finite for both Dirac and Weyl cases, irrespective of the presence and broken time reversal symmetry. While with the large Weyl node separation, the extrinsic part of the OH response gradually approaches zero, indicating a distinct dependence on the Weyl node separation compared with intrinsic contribution. However, this behavior does not imply that the extrinsic response is independent of band geometry, Berry connection, and the orbital structure of the system. 

Moreover, the application of external energy strengthens the extrinsic OH conductivity as shown in Fig.~\ref{fig 1}(c) and (g). The plateau in OH conductivity, arising from the interband transitions in the region of $\hbar\omega =\pm 2 \varepsilon_f$, broadens with increasing external energy, providing an effective control in the extrinsic OH response. In Fig.~\ref{fig 1}(d), we further examine the temperature dependence of the extrinsic OH conductivity. Here, increasing temperature leads to a broadening of the probability distribution, and reducing the occupation difference between the states near Fermi energy. Consequently, it results in a suppression of the extrinsic OH conductivity. It is worthwhile to note that the chiral index $\zeta$, corresponding to the two opposite chiralities for the different Weyl nodes does not change the behavior of $\sigma_{yx}^{ext}$. The overall extrinsic OH conductivity obtains a factor of two, due to the identical contribution from the two Weyl nodes with opposite chiralities. 
Furthermore, the contribution within the first Born approximation $\sigma_{yx}^{BA}$ to the extrinsic OH conductivity is negligible compared to the skew scattering beyond Born approximation $\sigma_{yx}^{BBA}$ from $f_{1E}^{\text{sk}}$, i.e., $\sigma_{yx}^{BBA} \approx 10^5\sigma_{yx}^{BA}$. Further, the details of the $\sigma_{yx}^{BA}$ contribution to extrinsic OH conductivity are provided in Appendix~\ref{App:B}.

To highlight the importance of extrinsic contribution over the intrinsic to the total OH conductivity, we compare the magnitudes of $\sigma_{yx}^{ext}$ with the existing $\sigma_{yx}^{int}$ for Weyl semimetal~\cite{Joy_NJP2026}. For a weak disorder potential, $U_0=30$ meV\AA$^3$, the extrinsic contribution yields a magnitude nearly 2.59 ($e/2\pi$) at a Fermi energy of 30 meV with Weyl node separation 0.05 \AA$^{-1}$ and applied energy with a value $\hbar\omega=60$meV.
While the intrinsic OH conductivity contributes 0.2986 $e/2\pi$ to the total OH conductivity at the same parameters, i.e., $\sigma_{yx}^{ext}\approx10\,\sigma_{yx}^{int}$. Further, the numerical estimation at different parameters is provided in Appendix~\ref{app:D}.

The OH response can exist in systems that preserve both the space inversion and time reversal symmetries as well as in the systems where either or both symmetries are broken, provided the relevant OAM matrix elements are finite~\cite{Kontani_PRL2008, Go_PRL2018, Busch_PRR2023, Sahu_PRB2021, Bhowal_PRB2021, Joy_PRB2025}. Under time-reversal symmetry, both the OAM operator and the velocity change sign, leaving the OHC operator invariant. In contrast, under spatial inversion, the OAM remains unchanged while the velocity changes sign, which makes the OHC operator odd. Nevertheless, the symmetry of the OH conductivity is determined by the transformation of the field-induced density matrix correction, in addition to the orbital current operator. Here, we considered a Weyl semimetal that preserves inversion symmetry but breaks time reversal. 

These above calculations are performed over the range of parameter chosen within the weak disorder limit and in the valid interband transitions. The disorder potential strength is varied from 0 to 1 eV\AA$^3$, which satisfy the disorder broadening remains small as compared to Fermi energy $\hbar/\tau\ll|\varepsilon_f|$. While the Fermi energy is varied within the interval $-\hbar\omega/2\leq\varepsilon_f\leq\hbar\omega/2$, where the external energy $\hbar\omega$ ranges from 10 to 100 meV. Further, these extrinsic mechanisms are not restricted to Weyl semimetals but can be more broadly applicable, including topologically trivial multiband systems~\cite{Choi_N2023, Veneri_PRL2025, Peng_NM2025}. Furthermore, for the case of a strong disorder strength, the perturbative expansion used here will not remain valid as the Bloch band picture breaks down and the Fermi surface as well as the topological features become ill-defined (not sharply defined). Hence for the higher-order or stronger potential one has to go beyond the perturbative expansion limit.

\section{Experimental relevance}
The experimental relevance of disorder-induced OH conductivity in time-reversal symmetry broken Weyl semimetals can be understood by relating the parameters considered in our calculations to experimentally controllable quantities. The OH response due to the flow of the OAM can be detected by the magneto-optic techniques, such as the magneto-optical Kerr effect (MOKE)~\cite{Choi_N2023, Jo_PRB2018}. The impurity potential strength can be tuned by selecting different dopants or defect types, enabling experimental verification of the disorder scaling of the OH conductivity~\cite{Nagaosa_RMP2010}, which is characteristic of skew scattering~\cite{Mankovsky_PRB2024, Du_NC2019, Messica_PRB2023}.
The dominance of extrinsic contributions can be identified by systematically varying impurity concentration while keeping the band structure intact. Furthermore, the predicted frequency-dependent response and the emergence of a stable conductivity plateau within the region $|\varepsilon_f|<\hbar\omega/2$ can be examined using terahertz or infrared optical conductivity measurements, with the Fermi energy tuned through chemical doping and gate controlled methods. 
The suppression of extrinsic OH conductivity with increasing Weyl node separation suggests that material platforms with strain-controlled Weyl node displacement~\cite{Xu_JPD2019} offer an experimental handle to distinguish disorder-driven OHEs from intrinsic topological contributions. To distinguish the OH response arising from Born and beyond Born approximations, systematically measure the OH conductivity across samples with different disorder strengths and compare the experimental results with theoretical predictions. Taken together, these features, such as disorder tunable magnitude, frequency-controlled plateau width and the suppression due to node separation, establish a clear and experimentally distinct extrinsic OH response in Weyl semimetals. We can consider the magnetic Weyl semimetals such as $\text{Co}_3\text{Sn}_2\text{S}_2$ and $\text{Mn}_3\text{Sn}$ as one of the suitable candidates for this investigation~\cite{ozawa_arxiv2026}. The system has the Weyl node separation in the range of $10^{-2}$ to $10^{-1}$ \AA$^{-1}$, which is under the first Brillouin zone~\cite{Bulmash_PRB2014, Li_PRB2021, Wang_PRL2016}. These systems can provide an OH conductivity in the meV region of the Fermi energy.

\section{Summary}
We have theoretically investigated the disorder-induced OH conductivity in three-dimensional Weyl semimetals with broken time-reversal symmetry using a quantum kinetic framework. Focusing on extrinsic mechanisms, we systematically analysed the contributions arising from side-jump and skew-scattering processes in the presence of impurity disorder. Our results demonstrate that the extrinsic OH conductivity exhibits a strong dependence on disorder strength, Fermi energy, external driving frequency, and Weyl node separation. We find that skew scattering provides the dominant contribution to the extrinsic OH conductivity, characterised by its linear dependence on the impurity potential. Furthermore, the tuning of the extrinsic OH conductivity by the parameters such as disorder potential, applied energy, and Weyl node separation, which elevates the magnitude of OH conductivity. Our findings establish three-dimensional Weyl semimetals as a promising platform for studying disorder-controlled orbital transport and provide new insights into the interplay between impurity scattering, band topology, and OAM dynamics. These results advance the theoretical understanding of extrinsic orbital Hall effects and open paths for tuning orbital currents in three-dimensional topological semimetals through disorder engineering.

\section*{Acknowledgment}
This work is financially supported by the Anusandhan National Research Foundation under project number SUR/2022/000289.
\section*{Data Availability Statement} All data that support the findings of this study are included within the article.
\onecolumngrid
\appendix
\section{Averages of disorder potential}\label{App:C}
Considering the short-range weak disorder potential which is randomly distributed $U(\bm r)=\sum_i U_0\delta(\bm r-\bm r_i)$, with impurity potential strength $U_0$ and the position of the $i^{\text{th}}$ impurity as $\bm r_i$. The disorder averages required for the collision integrals are evaluated by averaging over the impurity positions. For the first-order average, we obtain
\begin{equation}
    \langle U_{\bm q\bm q'}^{mm'}\rangle = \frac{U_0}{\mathcal{V}} \langle u_{\bm q}^m|u_{\bm q}^{m'}\rangle \langle \sum_i e^{i(\bm q' - \bm q)\cdot \bm r_i}\rangle.
\end{equation}
This yields average proportional to $\delta (\bm q - \bm q')$ which is diagonal in momentum space and vanishes for $\bm q \neq \bm q'$ and does not contribute to the collision integral. \\
For the second-order average, we have
\begin{equation}
    \langle U_{\bm q\bm q'}^{mm'}U_{\bm q'\bm q''}^{m'm''}\rangle= \frac{U_0^2}{\mathcal{V}^2}\langle u_{\bm q}^{m}|u_{\bm q'}^{m'}\rangle\langle u_{\bm q'}^{m'}|u_{\bm q''}^{m''}\rangle\bigg\langle\sum_{i,j}e^{i(\bm q'-\bm q)\cdot\bm r_i}e^{i(\bm q''-\bm q')\cdot\bm r_j}\bigg\rangle.
\end{equation} 
Here, the summation leads to two different cases with $i=j$ and $i \neq j$. The case $i = j$ refers to the scattering from same impurity which brings the average of phase factors as
\begin{equation}
    \bigg\langle\sum_{i}e^{i(\bm q'-\bm q)\cdot\bm r_i}e^{i(\bm q''-\bm q')\cdot\bm r_i}\bigg\rangle = \bigg\langle \sum_{i} e^{i(\bm q''-\bm q)\cdot\bm r_i}\bigg\rangle. 
\end{equation}
Here, the average producing the momentum conserving factor as $\delta (\bm q'' - \bm q)$. Therefore, we obtain 
\begin{equation}
    \langle U_{\bm q\bm q'}^{mm'}U_{\bm q'\bm q''}^{m'm''}\rangle=\frac{n_i U_0^2}{\mathcal{V}}\langle u_{\bm q}^{m}|u_{\bm q'}^{m'}\rangle\langle u_{\bm q'}^{m'}|u_{\bm q''}^{m''}\rangle\delta(\bm q''-\bm q),
\end{equation}
where $n_i$ is the impurity density. 
The other $i \neq j$ case refers to the scattering from two different impurities. Due to statistically independent impurity positions here, the overall average can be taken as product of first-order averages. Therefore, this case does not contribute to the connected momentum relaxing collision integral. \\
For the third-order average in the disorder potential, we have
\begin{align} \nonumber
  \langle U_{\bm q\bm q'}^{mm'}U_{\bm q'\bm q''}^{m'm''}U_{\bm q''\bm q'''}^{m''m'''}\rangle
  &=
  \frac{U_0^3}{\mathcal{V}^3}\langle u_{\bm q}^{m}|u_{\bm q'}^{m'}\rangle\langle u_{\bm q'}^{m'}|u_{\bm q''}^{m''}\rangle\langle u_{\bm q''}^{m''}|u_{\bm q'''}^{m'''}\rangle\\
  & \times \bigg\langle\sum_{i,j,k}e^{i(\bm q'-\bm q)\cdot\bm r_i}e^{i(\bm q''-\bm q')\cdot r_j}e^{i(\bm q'''-\bm q'')\cdot r_k}\bigg\rangle.
\end{align}
With scatterings from same impurity, $i = j = k$ which brings the phase factor average as
\begin{align} \nonumber
   \bigg\langle\sum_{i}e^{i(\bm q'-\bm q)\cdot\bm r_i}e^{i(\bm q''-\bm q')\cdot r_i}e^{i(\bm q'''-\bm q'')\cdot r_i}\bigg\rangle
   = \bigg\langle\sum_{i}e^{i(\bm q'''-\bm q)\cdot\bm r_i}\bigg\rangle.
\end{align}
This leads to the overall third order disorder average in the form
\begin{align}
  \langle U_{\bm q\bm q'}^{mm'}U_{\bm q'\bm q''}^{m'm''}U_{\bm q''\bm q'''}^{m''m'''}\rangle&=\frac{n_iU_0^3}{\mathcal{V}^2}\langle u_{\bm q}^{m}|u_{\bm q'}^{m'}\rangle\langle u_{\bm q'}^{m'}|u_{\bm q''}^{m''}\rangle\langle u_{\bm q''}^{m''}|u_{\bm q'''}^{m'''}\rangle\delta(\bm q'''-\bm q).
\end{align}
Further, the scattering from different impurity will not contribute to the momentum relaxing scattering term due to the statistically independent assumption for the impurity positions.

\section{Side-jump calculation}\label{App:A}
We present the detailed expressions for the collision integral by explicitly separating the diagonal and off-diagonal components of $\mathcal{J}_{\bm{q}}$ in band basis. We begin with the collision integral in Eq.~\eqref{Eq.4}, which is obtained for a multiband model within the first Born approximation, and is expressed as 
\begin{equation}
\mathcal{J}^{mp}_{\bm{q}}=\frac{i}{\hbar}\sum_{n, l}\sum_{{\bm q}'}\bigg\{\frac{\langle U_{\bm{qq'}}^{mn}U_{\bm{q'q}}^{lp}\rangle f_{E,\bm{q'}}^{nl}}{\varepsilon_{\bm{q'}}^{l}-\varepsilon_{\bm{q}}^{p}-i\eta}-\frac{\langle U_{\bm{qq'}}^{mn}U_{\bm{q'q}}^{nl}\rangle f_{E,\bm{q}}^{lp}}{\varepsilon_{\bm{q'}}^{n}-\varepsilon_{\bm{q}}^{l}-i\eta}+\frac{\langle U_{\bm{qq'}}^{mn}U_{\bm{q'q}}^{lp}\rangle f_{E,\bm{q'}}^{nl}}{\varepsilon_{\bm{q}}^{m}-\varepsilon_{\bm{q'}}^{n}-i\eta}-\frac{f_{E,\bm{q}}^{mn}\langle U_{\bm{qq'}}^{nl}U_{\bm{q'q}}^{lp}\rangle}{\varepsilon_{\bm{q}}^{n}-\varepsilon_{\bm{q'}}^{l}-i\eta}\bigg\},
\label{A:1}
\end{equation}
where the $m,n,l,p$ band indices, $U_{\bm{qq}'}^{mn}$ represents the disorder matrix, and $f_{E,{\bm q}}^{mn}$ is the field corrected density matrix. To identify the relevant contributions from the two-band model, the collision integral naturally separates into diagonal and off-diagonal components. Setting $m=p$ in the Eq.~\eqref{A:1}, the total diagonal contribution of the collision integral, 
\begin{equation}\begin{aligned}
\mathcal{J}^{mm}_{\bm{q}}=\frac{i}{\hbar}\sum_{{\bm q}'}\bigg\{&\frac{\langle U_{\bm{qq'}}^{mm}U_{\bm{q'q}}^{mm}\rangle f_{E,\bm{q'}}^{mm}}{\varepsilon_{\bm{q'}}^{m}-\varepsilon_{\bm{q}}^{m}-i\eta}-\frac{\langle U_{\bm{qq'}}^{mm}U_{\bm{q'q}}^{mm}\rangle f_{E,\bm{q}}^{mm}}{\varepsilon_{\bm{q'}}^{m}-\varepsilon_{\bm{q}}^{m}-i\eta}+\frac{\langle U_{\bm{qq'}}^{mm}U_{\bm{q'q}}^{mm}\rangle f_{E,\bm{q'}}^{mm}}{\varepsilon_{\bm{q}}^{m}-\varepsilon_{\bm{q'}}^{m}-i\eta}-\frac{f_{E,\bm{q}}^{mm}\langle U_{\bm{qq'}}^{mm}U_{\bm{q'q}}^{mm}\rangle}{\varepsilon_{\bm{q}}^{m}-\varepsilon_{\bm{q'}}^{m}-i\eta}\\&+\frac{\langle U_{\bm{qq'}}^{mp}U_{\bm{q'q}}^{mm}\rangle f_{E,\bm{q'}}^{pm}}{\varepsilon_{\bm{q'}}^{m}-\varepsilon_{\bm{q}}^{m}-i\eta}-\frac{\langle U_{\bm{qq'}}^{mp}U_{\bm{q'q}}^{pm}\rangle f_{E,\bm{q}}^{mm}}{\varepsilon_{\bm{q'}}^{p}-\varepsilon_{\bm{q}}^{m}-i\eta}+\frac{\langle U_{\bm{qq'}}^{mp}U_{\bm{q'q}}^{mm}\rangle f_{E,\bm{q'}}^{pm}}{\varepsilon_{\bm{q}}^{m}-\varepsilon_{\bm{q'}}^{p}-i\eta}-\frac{f_{E,\bm{q}}^{mp}\langle U_{\bm{qq'}}^{pm}U_{\bm{q'q}}^{mm}\rangle}{\varepsilon_{\bm{q}}^{p}-\varepsilon_{\bm{q'}}^{m}-i\eta}\\&+\frac{\langle U_{\bm{qq'}}^{mm}U_{\bm{q'q}}^{pm}\rangle f_{E,\bm{q'}}^{mp}}{\varepsilon_{\bm{q'}}^{p}-\varepsilon_{\bm{q}}^{m}-i\eta}-\frac{\langle U_{\bm{qq'}}^{mm}U_{\bm{q'q}}^{mp}\rangle f_{E,\bm{q}}^{pm}}{\varepsilon_{\bm{q'}}^{m}-\varepsilon_{\bm{q}}^{p}-i\eta}+\frac{\langle U_{\bm{qq'}}^{mm}U_{\bm{q'q}}^{pm}\rangle f_{E,\bm{q'}}^{mp}}{\varepsilon_{\bm{q}}^{m}-\varepsilon_{\bm{q'}}^{m}-i\eta}-\frac{f_{E,\bm{q}}^{mm}\langle U_{\bm{qq'}}^{ml}U_{\bm{q'q}}^{pm}\rangle}{\varepsilon_{\bm{q}}^{m}-\varepsilon_{\bm{q'}}^{p}-i\eta}\\&+\frac{\langle U_{\bm{qq'}}^{mp}U_{\bm{q'q}}^{pm}\rangle f_{E,\bm{q'}}^{pp}}{\varepsilon_{\bm{q'}}^{p}-\varepsilon_{\bm{q}}^{m}-i\eta}-\frac{\langle U_{\bm{qq'}}^{mp}U_{\bm{q'q}}^{pp}\rangle f_{E,\bm{q}}^{pm}}{\varepsilon_{\bm{q'}}^{p}-\varepsilon_{\bm{q}}^{p}-i\eta}+\frac{\langle U_{\bm{qq'}}^{mp}U_{\bm{q'q}}^{pm}\rangle f_{E,\bm{q'}}^{pp}}{\varepsilon_{\bm{q}}^{m}-\varepsilon_{\bm{q'}}^{p}-i\eta}-\frac{f_{E,\bm{q}}^{mp}\langle U_{\bm{qq'}}^{pp}U_{\bm{q'q}}^{pm}\rangle}{\varepsilon_{\bm{q}}^{p}-\varepsilon_{\bm{q'}}^{p}-i\eta}\bigg\}.
\label{A:2}
\end{aligned}\end{equation}
Here, we separate the terms for the diagonal distribution of the density matrix in the diagonal collision integral,
\begin{equation}\begin{aligned}
\mathcal{J}^{mm}_{\bm{q}}(f_{E,\bm q}^{mm})=\frac{2\pi}{\hbar}\sum_{{\bm q}'}&\bigg\{\langle U_{\bm{qq'}}^{mm}U_{\bm{q'q}}^{mm}\rangle \bigg(f_{E,\bm{q'}}^{mm}-f_{E,\bm{q}}^{mm}\bigg)\delta(\varepsilon_{\bm{q}}^{m}-\varepsilon_{\bm{q'}}^{m})\\&+\langle U_{\bm{qq'}}^{mp}U_{\bm{q'q}}^{pm}\rangle \bigg(f_{E,\bm{q'}}^{pp}-f_{E,\bm{q}}^{mm}\bigg)\delta(\varepsilon_{\bm{q'}}^{p}-\varepsilon_{\bm{q}}^{m})\bigg\}.
\label{A:3}
\end{aligned}\end{equation}
Since, we do not have prior information about the diagonal component of the density matrix $f_{E,{\bm q}}^{mm}$, thus we treat the above equation within the relaxation time approximation as $f_{E,{\bm q}}^{mm}/\tau_{\bm q}^{m}$. The remaining terms are proportional to the off-diagonal density matrix elements,
\begin{equation}\begin{aligned}
\mathcal{J}^{mm}_{\bm{q}}(f_{E,\bm q}^{mp})=\frac{i}{\hbar}\sum_{{\bm q}'}\bigg\{&\bigg(\langle U_{\bm{qq'}}^{mp}U_{\bm{q'q}}^{mm}\rangle f_{E,\bm{q'}}^{pm}+\langle U_{\bm{qq'}}^{mm}U_{\bm{q'q}}^{pm}\rangle f_{E,\bm{q'}}^{mp}\bigg)\bigg(\delta(\varepsilon_{\bm{q'}}^{m}-\varepsilon_{\bm{q}}^{m})+\delta(\varepsilon_{\bm{q}}^{m}-\varepsilon_{\bm{q'}}^{p})\bigg)\\&-\bigg(f_{E,\bm{q}}^{mp}\langle U_{\bm{qq'}}^{pm}U_{\bm{q'q}}^{mm}\rangle+f_{E,\bm{q}}^{pm}\langle U_{\bm{qq'}}^{mm}U_{\bm{q'q}}^{mp}\rangle\bigg)\delta(\varepsilon_{\bm{q}}^{p}-\varepsilon_{\bm{q'}}^{m})\\&-\bigg(f_{E,\bm{q}}^{mp}\langle U_{\bm{qq'}}^{pp}U_{\bm{q'q}}^{pm}\rangle-f_{E,\bm{q}}^{pm}\langle U_{\bm{qq'}}^{mp}U_{\bm{q'q}}^{pp}\rangle\bigg)\delta(\varepsilon_{\bm{q}}^{p}-\varepsilon_{\bm{q'}}^{p})\bigg\}.
\label{A:4}
\end{aligned}\end{equation}
Further choosing $m\neq p$, the complete off-diagonal collision integral can be expressed as follows,
\begin{equation}\begin{aligned}
\mathcal{J}^{mp}_{\bm{q}}=\frac{i}{\hbar}\sum_{{\bm q}'}\bigg\{&\frac{\langle U_{\bm{qq'}}^{mm}U_{\bm{q'q}}^{mp}\rangle f_{E,\bm{q'}}^{mm}}{\varepsilon_{\bm{q'}}^{m}-\varepsilon_{\bm{q}}^{p}-i\eta}-\frac{\langle U_{\bm{qq'}}^{mm}U_{\bm{q'q}}^{mm}\rangle f_{E,\bm{q}}^{mp}}{\varepsilon_{\bm{q'}}^{m}-\varepsilon_{\bm{q}}^{m}-i\eta}+\frac{\langle U_{\bm{qq'}}^{mm}U_{\bm{q'q}}^{mp}\rangle f_{E,\bm{q'}}^{mm}}{\varepsilon_{\bm{q}}^{m}-\varepsilon_{\bm{q'}}^{m}-i\eta}-\frac{f_{E,\bm{q}}^{mm}\langle U_{\bm{qq'}}^{mm}U_{\bm{q'q}}^{mp}\rangle}{\varepsilon_{\bm{q}}^{m}-\varepsilon_{\bm{q'}}^{m}-i\eta}\\&+
\frac{\langle U_{\bm{qq'}}^{mp}U_{\bm{q'q}}^{mp}\rangle f_{E,\bm{q'}}^{pm}}{\varepsilon_{\bm{q'}}^{m}-\varepsilon_{\bm{q}}^{p}-i\eta}-\frac{\langle U_{\bm{qq'}}^{mp}U_{\bm{q'q}}^{pm}\rangle f_{E,\bm{q}}^{mp}}{\varepsilon_{\bm{q'}}^{p}-\varepsilon_{\bm{q}}^{m}-i\eta}+\frac{\langle U_{\bm{qq'}}^{mp}U_{\bm{q'q}}^{mp}\rangle f_{E,\bm{q'}}^{pm}}{\varepsilon_{\bm{q}}^{m}-\varepsilon_{\bm{q'}}^{p}-i\eta}-\frac{f_{E,\bm{q}}^{mp}\langle U_{\bm{qq'}}^{pm}U_{\bm{q'q}}^{mp}\rangle}{\varepsilon_{\bm{q}}^{p}-\varepsilon_{\bm{q'}}^{m}-i\eta}\\&+
\frac{\langle U_{\bm{qq'}}^{mm}U_{\bm{q'q}}^{pp}\rangle f_{E,\bm{q'}}^{mp}}{\varepsilon_{\bm{q'}}^{p}-\varepsilon_{\bm{q}}^{p}-i\eta}-\frac{\langle U_{\bm{qq'}}^{mm}U_{\bm{q'q}}^{mp}\rangle f_{E,\bm{q}}^{pp}}{\varepsilon_{\bm{q'}}^{m}-\varepsilon_{\bm{q}}^{p}-i\eta}+\frac{\langle U_{\bm{qq'}}^{mm}U_{\bm{q'q}}^{pp}\rangle f_{E,\bm{q'}}^{mp}}{\varepsilon_{\bm{q}}^{m}-\varepsilon_{\bm{q'}}^{m}-i\eta}-\frac{f_{E,\bm{q}}^{mm}\langle U_{\bm{qq'}}^{mp}U_{\bm{q'q}}^{pp}\rangle}{\varepsilon_{\bm{q}}^{m}-\varepsilon_{\bm{q'}}^{p}-i\eta}\\&+
\frac{\langle U_{\bm{qq'}}^{mp}U_{\bm{q'q}}^{pp}\rangle f_{E,\bm{q'}}^{pp}}{\varepsilon_{\bm{q'}}^{p}-\varepsilon_{\bm{q}}^{p}-i\eta}-\frac{\langle U_{\bm{qq'}}^{mp}U_{\bm{q'q}}^{pp}\rangle f_{E,\bm{q}}^{pp}}{\varepsilon_{\bm{q'}}^{p}-\varepsilon_{\bm{q}}^{p}-i\eta}+\frac{\langle U_{\bm{qq'}}^{mp}U_{\bm{q'q}}^{pp}\rangle f_{E,\bm{q'}}^{pp}}{\varepsilon_{\bm{q}}^{m}-\varepsilon_{\bm{q'}}^{p}-i\eta}-\frac{f_{E,\bm{q}}^{mp}\langle U_{\bm{qq'}}^{pp}U_{\bm{q'q}}^{pp}\rangle}{\varepsilon_{\bm{q}}^{p}-\varepsilon_{\bm{q'}}^{p}-i\eta}\bigg\}.
\label{A:5}
\end{aligned}\end{equation}
This expression also originates from the diagonal and off-diagonal contributions, which can be separated as shown below.
\begin{equation}\begin{aligned}
\mathcal{J}^{mp}_{\bm{q}}(f_{E,\bm q}^{mm})=\frac{\pi}{\hbar}\sum_{{\bm q}'}&\bigg\{\langle U_{\bm{qq'}}^{mm}U_{\bm{q'q}}^{mp}\rangle\bigg(f_{E,\bm{q'}}^{mm}-f_{E,\bm{q}}^{pp}\bigg)\delta(\varepsilon_{\bm{q'}}^{m}-\varepsilon_{\bm{q}}^{p})\\&+\langle U_{\bm{qq'}}^{mm}U_{\bm{q'q}}^{mp}\rangle\bigg(f_{E,\bm{q'}}^{mm}-f_{E,\bm{q}}^{mm}\bigg)\delta(\varepsilon_{\bm{q}}^{m}-\varepsilon_{\bm{q'}}^{m})\\&+\langle U_{\bm{qq'}}^{mp}U_{\bm{q'q}}^{pp}\rangle \bigg(f_{E,\bm{q'}}^{pp}-f_{E,\bm{q}}^{pp}\bigg)\delta(\varepsilon_{\bm{q'}}^{p}-\varepsilon_{\bm{q}}^{p})\\&+\langle U_{\bm{qq'}}^{mp}U_{\bm{q'q}}^{pp}\rangle\bigg( f_{E,\bm{q'}}^{pp}-f_{E,\bm{q}}^{mm}\bigg)\delta(\varepsilon_{\bm{q}}^{m}-\varepsilon_{\bm{q'}}^{p})\bigg\}.
\label{A:6}
\end{aligned}\end{equation}
The remaining contribution from the off-diagonal density matrix for the off-diagonal collision integral is
\begin{equation}\begin{aligned}
\mathcal{J}^{mp}_{\bm{q}}(f_{E,\bm q}^{mp})=\frac{\pi}{\hbar}\sum_{{\bm q}'}&\bigg\{\bigg(\langle U_{\bm{qq'}}^{mp}U_{\bm{q'q}}^{mp}\rangle f_{E,\bm{q'}}^{pm}-f_{E,\bm{q}}^{mp}\langle U_{\bm{qq'}}^{pm}U_{\bm{q'q}}^{mp}\rangle\bigg)\delta(\varepsilon_{\bm{q}}^{p}-\varepsilon_{\bm{q'}}^{m})
\\&+\bigg(\langle U_{\bm{qq'}}^{mp}U_{\bm{q'q}}^{mp}\rangle f_{E,\bm{q'}}^{pm}-f_{E,\bm{q}}^{mp}\langle U_{\bm{qq'}}^{mp}U_{\bm{q'q}}^{pm}\rangle\bigg) \delta(\varepsilon_{\bm{q'}}^{p}-\varepsilon_{\bm{q}}^{m})
\\&+\bigg(f_{E,\bm{q'}}^{mp}\langle U_{\bm{qq'}}^{mm}U_{\bm{q'q}}^{pp}\rangle -f_{E,\bm{q}}^{mp}\langle U_{\bm{qq'}}^{pp}U_{\bm{q'q}}^{pp}\rangle\bigg)\delta(\varepsilon_{\bm{q}}^{p}-\varepsilon_{\bm{q'}}^{p})\\&+\bigg(f_{E,\bm{q'}}^{mp}\langle U_{\bm{qq'}}^{mm}U_{\bm{q'q}}^{pp}\rangle-f_{E,\bm{q}}^{mp}\langle U_{\bm{qq'}}^{mm}U_{\bm{q'q}}^{mm}\rangle\bigg)\delta(\varepsilon_{\bm{q'}}^{m}-\varepsilon_{\bm{q}}^{m})\bigg\},
\label{A:7}
\end{aligned}\end{equation}
The above equation is treated within the relaxation time approximation as $f_{E,{\bm q}}^{mp}/\tau$, considering $\tau$ as a constant. \\ These expressions are simplified using the short-range disorder model to obtain the side-jump contributions to the OH conductivity. 
\section{Orbital Hall Conductivity in the First Born Approximation}\label{App:B}
\begin{figure}[t]
    \centering
    \includegraphics[width=1\linewidth]{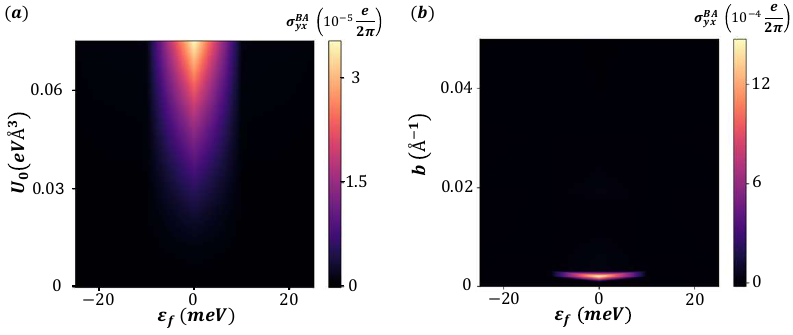}
    \caption{The plots depict the relation between total OH conductivity arising from the first Born approximation $(\sigma_{yx}^{BA})$ on Fermi energy $(\varepsilon_f)$. (a) and (b) shows the variation of $\sigma_{yx}^{BA}$ with disorder potential ($U_0$) and Weyl node separation ($b$) with an applied energy ($\hbar\omega=20$meV) at the temperature $T=1$K. The remaining parameters in each panel are fixed at $b=0.04\,\text{\AA}^{-1}$, and $U_0=50\,\text{meV\AA}^3$.}
    \label{fig:APP.C}
\end{figure}

In Fig.~\ref{fig:APP.C} (a) and (b), the side jump and the skew scattering contributions obtained within the first Born approximation ($\sigma_{yx}^{BA}$) are illustrated as a function of Fermi energy for different disorder potentials and Weyl node separations. As we mentioned in the main text, the contribution from the first Born approximation is negligible compared to the beyond Born approximation ($\sigma_{yx}^{BBA}$) in the total extrinsic OH conductivity. 
%
The results clearly show that the magnitudes of the two contributions differ by five orders, i.e., $\sigma_{yx}^{BBA}\approx10^5\sigma_{yx}^{BA}$. Therefore, the first Born approximation provides only a negligible correction to the total extrinsic OH response.

\section{Numerical Comparison of different contributions to OH response}
\label{app:D}
The numerical comparison of the three mechanisms that contribute to the overall OH conductivity is shown in the Table~\ref{tab} with different disorder potential ($U_0$), Weyl node parameter ($b$) and Fermi energy ($\varepsilon_f$). 
%
\begin{table*}[ht]
\caption{Numerical comparison of magnitudes of intrinsic, side-jump and skew scattering contributions to the orbital Hall conductivity on the disorder strength $U_0$ at temperature $T=1$~K.}
\label{tab}
\begin{ruledtabular}
\begin{tabular*}{\textwidth}{@{\extracolsep{\fill}}cccccc}
$U_0$ (meV \AA$^3$) & 
$b$ (\AA$^{-1}$) & 
$\varepsilon_f$ (meV) & 
$\sigma_{yx}^{\mathrm{int}}(e/2\pi)$ & 
$\sigma_{yx}^{\mathrm{sj}}(e/2\pi)$ & 
$\sigma_{yx}^{\mathrm{sk}}(e/2\pi)$ \\
\multicolumn{3}{c}{}&
\multicolumn{3}{c}{}\\
\hline
10 & 0.01 & 20 & 0.00601 & $\sim 4.1\times10^{-7}$   & $\sim 0.184$   \\
10 & 0.05 & 20 & 0.277   & $\sim 1.286\times10^{-7}$ & $\sim 0.00063$ \\
10 & 0.05 & 30 & 0.2986  & $\sim 5.97\times10^{-7}$  & $\sim 0.0948$  \\
30 & 0.05 & 30 & 0.2986  & $\sim 5.37\times10^{-6}$  & $\sim 2.594$   \\
50 & 0.05 & 30 & 0.2986  & $\sim 1.4\times10^{-6}$   & $\sim 12.1205$ \\
\end{tabular*}
\end{ruledtabular}
\end{table*}

\twocolumngrid
\bibliography{Ref,Ref2}

\end{document}